# Exploring the Utility of MALDI-TOF Mass Spectrometry and Antimicrobial Resistance in Hospital Outbreak Detection

**Chang Liu, MS[1], Jieshi Chen, MS[1], Alexander J. Sundermann, DrPH[2,3,4], Kathleen Shutt, MS [2,3], Marissa P. Griffith, BS [2,3], Lora Lee Pless, PhD [2,3], Lee H. Harrison, MD[2,3,4], Artur W. Dubrawski, PhD[1].**

**[1]Auton Lab, Carnegie Mellon University, Pittsburgh, Pennsylvania, USA;**
**[2]Microbial Genomic Epidemiology Laboratory, Center for Genomic Epidemiology, University of Pittsburgh, Pittsburgh, Pennsylvania, USA;**
**[3]Division of Infectious Diseases, University of Pittsburgh School of Medicine, Pittsburgh, Pennsylvania, USA;**
**[4]Department of Epidemiology, School of Public Health, University of Pittsburgh, Pittsburgh, Pennsylvania, USA.**

**Abstract**

*Accurate and timely identification of hospital outbreak clusters is crucial for preventing the spread of infections that have epidemic potential. While assessing pathogen similarity through whole genome sequencing (WGS) is considered the gold standard for outbreak detection, its high cost and lengthy turnaround time preclude routine implementation in clinical laboratories. We explore the utility of two rapid and cost-effective alternatives to WGS, matrix-assisted laser desorption ionization-time of flight (MALDI-TOF) mass spectrometry and antimicrobial resistance (AR) patterns. We develop a machine learning framework that extracts informative representations from MALDI-TOF spectra and AR patterns for outbreak detection and explore their fusion. Through multi-species analyses, we demonstrate that in some cases MALDI-TOF and AR have the potential to reduce reliance on WGS, enabling more accessible and rapid outbreak surveillance.*

## Introduction

Disease outbreaks in hospitals can occur when pathogens spread unchecked among patients, medical equipment, and healthcare workers[1,2]. Identifying these outbreaks at their early stages is vital to preventing infections and limiting the resources needed to address widespread transmission[3]. To achieve this, healthcare institutions compile epidemiological data from the electronic health record and, separately, collect pathogen data from patient and environmental samples to detect and address potential outbreak clusters[4]. Consisting of samples grouped by pathogen similarity, these clusters are considered connected to recent transmission events[5,6], enabling the tracking of pathogen dissemination within hospitals[7].

Whole genome sequencing (WGS) stands as the gold standard for assessing pathogen similarity, currently providing the most information and highest discriminatory power[8]. The technology has been used to better understand the transmission of multiple pathogens, e.g., *Pseudomonas aeruginosa* (PSA)[9,10] and vancomycin-resistant *Enterococcus faecium* (VRE)[11,12]. Specifically, two cases belong to the same cluster if they are separated by single-nucleotide polymorphisms (SNPs) fewer than a specified threshold[13,14]. However, the cost and turnaround time required for WGS have prohibited its widespread deployment in clinical laboratories[15].

To attain cost-effective and rapid outbreak detection that can be deployed with ease, we explore two existing alternatives to WGS: matrix-assisted laser desorption ionization–time of flight (MALDI-TOF) mass spectrometry and antimicrobial resistance (AR) patterns. MALDI-TOF is a routine tool for identifying bacterial species in clinical laboratories[16,17] and has been suggested as a possible substitute for WGS in outbreak cluster detection due to its affordability and rapid turnaround time[18]. It produces intensity spectra based on mass-to-charge ratios (m/z) of microbial proteins, creating a distinctive "fingerprint" for each microorganism. Subsequently, these spectra undergo analysis to identify outbreak clusters[19,20]. MALDI-TOF has shown promise in investigating the epidemiology of outbreaks associated with VRE[18], *Streptococcus pneumoniae* (SPN)[21], and methicillin resistant *Staphylococcus aureus* (MRSA)[22]. However, it has also been shown that MALDI-TOF disagreed with WGS-defined ground truth clusters for VRE[23] and *Klebsiella pneumoniae* (KLP)[24] outbreaks. Nevertheless, such studies are often limited in scope by examining small-scale data of only one single pathogen at a time. Furthermore, without eliciting crucial, relevant knowledge embedded in the MALDI-TOF spectra, their raw form may not be sufficient to produce clustering results with optimal similarity to WGS.

Leveraging data already collected by clinical laboratories, outbreak detection through analysis of AR patterns is attractive for its minimal resource burden[25] and near real-time speed[26]. It has witnessed success in detecting outbreak clusters associated with *Shigella* spp.[25] and VRE[27]. However, while AR often yields high sensitivity in outbreak detection, researchers noted a lack of specificity in the identified clusters[28]. Nevertheless, there has been a lack of research that compares the utility of AR with other candidate outbreak detection methods, such as MALDI-TOF.

To fill these gaps, we propose a machine learning-based framework that extracts representations of MALDI-TOF and AR data tailored to hospital outbreak cluster detection. Using a comprehensive dataset, we conduct a multi-species analysis that compares clustering results of MALDI-TOF and AR with those of WGS. We evaluate the utility of MALDI-TOF, AR, and their synergy, in species-agnostic and species-specific outbreak detection settings. We also demonstrate that in some cases MALDI-TOF and AR could potentially act as cost-effective and rapid alternatives by reducing the need for WGS in outbreak detection.

**Methods**

*Data source*: Our dataset is de-identified and proprietary, obtained from a large non-profit research and academic hospital. It consists of 4921 isolates spanning 17 bacterial species (Table 1), collected from 10/2021 to 10/2024 duri ng routine surveillance and were used to identify active outbreaks. Each isolate contains a raw MALDI-TOF spectrum and a phenotypic antimicrobial resistance profile. For each pair of isolates under the same species, we collect their SNP distance derived from WGS experiments. Ground truth outbreak clusters are determined by performing hierarchical clustering (with complete linkage) on the complete SNP distance matrix with a cutoff distance of 15. This threshold reflects the standard practice adopted by the hospital infection control team that provided our proprietary data, serving as their routine reference point for detecting emerging outbreak clusters. In our experiments, we perform 4-fold cross-validation on isolates, grouped by clusters: we preserve the percentage of samples of each species across the data splits while ensuring that the isolates from one ground truth outbreak cluster will fall in only one split. In cr oss-validation experiments, only the SNP distances between isolates within the training splits are observable to the model.

**Table 1.** Species-wise dataset statistics. "# Clusters" represent the count of outbreak clusters with more than one isolate. "# Singletons" denote the number of isolates without forming any outbreak clusters with other isolates.

| Species | Abbreviation | # Isolates | # Clusters | # Singletons |
|---|---|---|---|---|
| *Acinetobacter baumannii* | ACIN | 137 | 8 | 105 |
| *Burkholderia cepaciae* | BC | 28 | 4 | 12 |
| *Citrobacter* | CB | 105 | 7 | 89 |
| *Enterobacter cloacae* | EB | 113 | 13 | 84 |
| *Escherichia coli* | EC | 303 | 26 | 240 |
| *Klebsiella oxytoca* | KLO | 45 | 2 | 40 |
| *Klebsiella pneumoniae* | KLP | 273 | 41 | 138 |
| *Staphylococcus aureus* | MRSA | 544 | 44 | 441 |
| *Mycobacterium* | MYC | 87 | 8 | 53 |
| *Proteus mirabilis* | PR | 683 | 53 | 549 |
| *Providencia* | PRV | 81 | 2 | 76 |
| *Pseudomonas aeruginosa* | PSA | 1543 | 176 | 987 |
| *Pseudomonas* | PSB | 66 | 2 | 62 |
| *Serratia marcescens* | SER | 364 | 40 | 255 |
| *Stenotrophomonas maltophilia* | STEN | 268 | 18 | 212 |
| *Vancomycin-resistant Enterococcus* | VRE | 270 | 22 | 137 |
| Multiple species (others) | MULT | 11 | 2 | 7 |
| Total | | 4921 | 468 | 3487 |

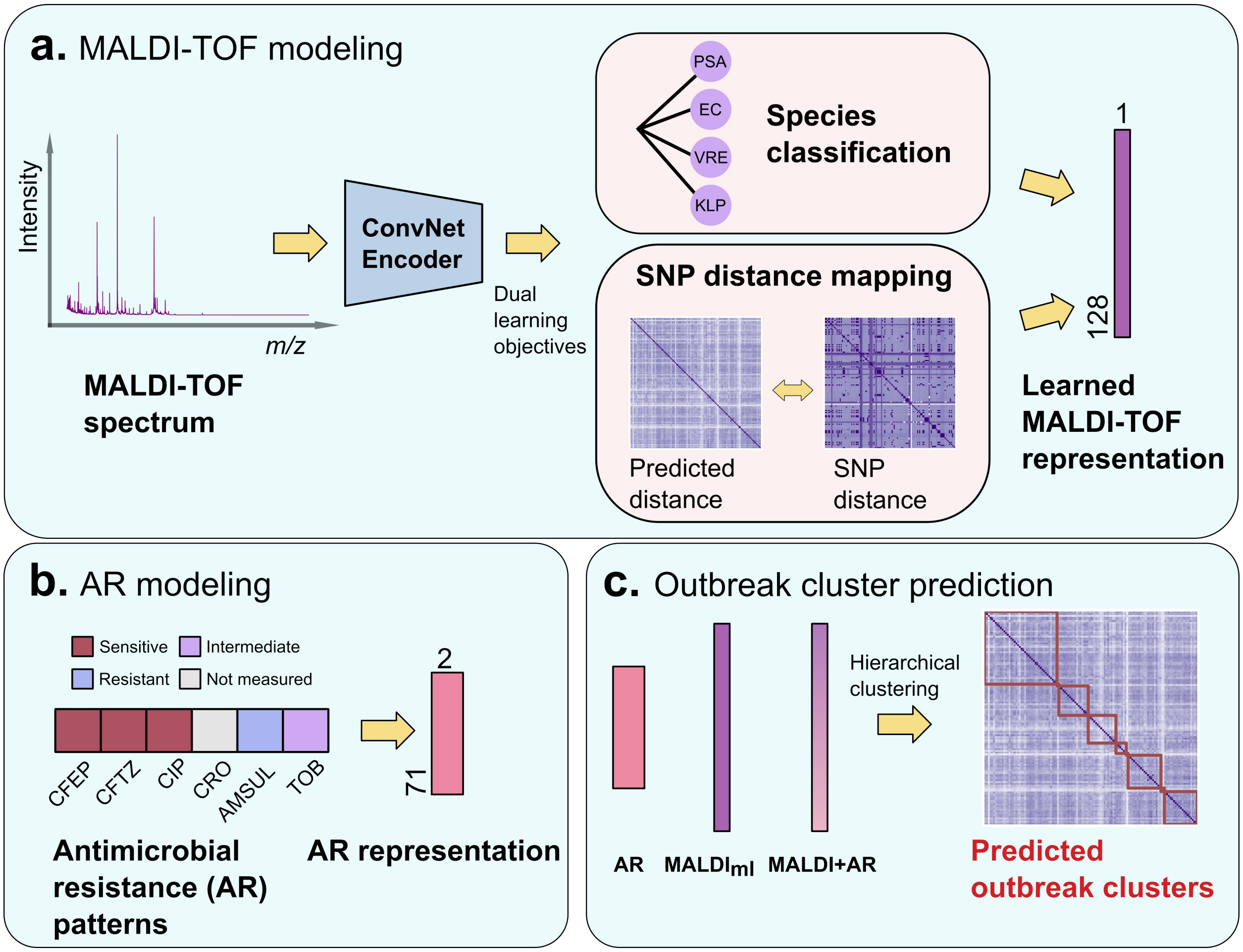

**Figure 1.** Overview of our framework. **a**. We adopt a machine learning-based approach to learn representations of MALDI-TOF spectra, using a 1D convolutional neural network with dual learning objectives. b. We model antimicr obial resistance (AR) patterns by separately encoding different outcomes. c. We employ hierarchical clustering on re presentations of AR, MALDI, and their combinations to obtain predicted outbreak clusters.

*Modeling MALDI-TOF spectra*: In the initial preprocessing step, the unprocessed spectral data—consisting of mass-to-charge ratio (*m*/*z*) and intensity pairs—undergo filtering to retain only those pairs with *m*/*z* values ranging from 2000 to 20000 Daltons. These filtered data are then discretized into 2000 uniformly distributed bins.

We then train a machine learning model that learns compact MALDI-TOF representations useful for outbreak detection (Figure 1a). The trained model aims to estimate similarities using only MALDI-TOF data without requiring external information, e.g., SNP distance or WGS results. As illustrated in Figure 1a, our framework entails two learning objectives: (1) identify the species of the isolate from the input MALDI-TOF spectrum, and (2) map the distances between the learned MALDI-TOF representations to the ground truth SNP distances.

We employ a 1D convolutional neural network architecture to encode the MALDI spectra (Figure 1a). The neural ne twork first encodes input spectra via stacked convolutional blocks (channels: 1→32→64→128), each containing two 3×3 convolutions with batch normalization. The encoded features are then projected to a 128-dimensional embedding space via 1×1 convolution and a fully connected layer. These embeddings are then fed into a classifier of two fully connected layers for species prediction and are simultaneously optimized for SNP distance mapping.

Formally, the learning objective/loss function is defined as:

$$\mathscr{L} = \lambda \mathrm{CE}(z,y) + \mu L_{\mathrm{SNP}},$$

where $z$ is the prediction probability for each species, $y$ stands for the ground truth species labels, $\mathrm{CE}(\cdot)$ is the cross-entropy loss function, and $L_{\mathrm{SNP}}$ is a SNP distance learning objective[29]. Specifically, for two isolates of the same species with SNP distance $d$ and learned MALDI-TOF features having Euclidian distance $\hat{d}$: if $d \leq t$ for a predefined threshold $t$, then $L_{\mathrm{SNP}} = \left(d - \hat{d}\right)^2$. If $d > t$, then $L_{\mathrm{SNP}} = \max\{0, t - \hat{d}\}^2$. Here, $t$ is set to 15, the cutoff distance when obtaining the ground truth clusters using SNP distances. This SNP distance learning objective imposes no penalty when both the feature distance and SNP distance exceed the SNP threshold, as they have no impact on the results of outbreak cluster detection but may cause the model to overfit to large SNP distances ($> 10^5$). In addition, $\lambda$ and $\mu$ are hyperparameters controlling the contribution of each loss component and are searched in the grid of $[0.1, 1, 10] \times [0.1, 1, 10]$. In our experiments the combination $(\lambda, \mu) = (10, 0.1)$ achieved the best clustering performance and was chosen as our hyperparameter configuration. We denote this model by "$\mathrm{MALDI}_{\mathrm{ml}}$."

To assess the importance of either learning objective, we conduct an ablation study by developing two models for comparison: $\mathrm{MALDI}_{\mathrm{sp}}$ and $\mathrm{MALDI}_{\mathrm{snp}}$, which retains only the species classification objective ($\mathrm{CE}(z,y)$) and the SNP distance mapping objective ($L_{\mathrm{SNP}}$), respectively. All models are also compared with the baseline, $\mathrm{MALDI}_{\mathrm{raw}}$, which represents binned MALDI-TOF spectra in its raw form without any learning.

*Modeling AR patterns*: The antimicrobial resistance of each isolate is measured against a subset of 71 antibiotics, with outcomes among {"Sensitive", "Intermediate", "Resistant", "Not measured"} (Figure 1b). We embed each outcome into a 2-dimensional vector: "Sensitive" → [1,0], "Intermediate" → [0.5, 0.5], "Resistant" → [0,1]. Note that we encode "Not measured" to [0,0] since it represents maximum uncertainty while being orthogonal to the other three outcomes. Hence, the AR pattern for each isolate is represented as a vector of dimension $71 \times 2$ and flattened into a 142-dimensional vector that captures the discrepancy and similarity between outcome categories.

*Fusing MALDI-TOF and AR representations*: We explore fusing the above MALDI-TOF (using $\mathrm{MALDI}_{\mathrm{ml}}$) and AR representations into a unified vector for outbreak detection. We first standardize the AR representations into zero-mean, unit-variance vectors. Then, we apply principal component analysis (PCA) to the standardized AR representations to retrieve the top 128 principal components and sum the MALDI and resulting AR representations, weighted by $\alpha$ and $1 - \alpha$, respectively. Here, $\alpha$ is a hyperparameter controlling the relative importance of the two representations. We searched $\alpha$ among $[0.25, 0.5, 0.75]$ and found that $\alpha = 0.25$ achieved the best downstream clustering performance. This combined method is denoted by "MALDI+AR."

*Detecting and evaluating outbreak clusters*: To identify output clusters from the above representations, we apply hierarchical clustering with Ward linkage, using a distance threshold (Figure 1c). To determine the optimal threshold, we select the distance that maximizes the *outbreak F1 score* (defined below) on the training data for each species and cross-validation fold.

We evaluate performance by reformulating clustering as a pairwise binary classification problem. Specifically:

- Positive class: Isolate pairs assigned to the same SNP-derived ground truth cluster
- Negative class: Isolate pairs assigned to different clusters

This framework allows us to define standard classification metrics:

- True Positives (TP): Pairs correctly predicted to be in the same cluster
- True Negatives (TN): Pairs correctly predicted to be in different clusters
- False Positives (FP): Pairs incorrectly predicted to be in the same cluster
- False Negatives (FN): Pairs incorrectly predicted to be in different clusters

To assess the utility of MALDI-TOF and AR from the clustering results, we employ both machine learning and clinical evaluation metrics. For machine learning metrics, we adopt normalized mutual information (NMI) and the adjusted rand index (ARI). For clinical metrics, we define *sensitivity* and *positive predictive value* (PPV) as follows:

$$\mathrm{Sensitivity} = \frac{\mathrm{TP}}{\mathrm{TP} + \mathrm{FN}},$$

$$\mathrm{PPV} = \frac{\mathrm{TP}}{\mathrm{TP} + \mathrm{FP}}.$$

Sensitivity and PPV reflect different facets of outbreak detection: sensitivity quantifies the proportion of ground truth clusters identified by our method, while PPV measures the proportion of predicted clusters that are confirmed by WGS ground truth.

To take both sensitivity and PPV into account for model selection, we define the *outbreak F1 score* as their harmonic average:

$$\mathrm{F1} = \frac{2 \times \mathrm{Sensitivity} \times \mathrm{PPV}}{\mathrm{Sensitivity} + \mathrm{PPV}} .$$

This metric is used to optimize the distance threshold for hierarchical clustering, ensuring that the resulting clusters achieve an appropriate balance between identifying true outbreaks and minimizing false associations.

**Results**

*Species-agnostic utility*: We first evaluated the overall outbreak detection and clustering performance across all species in the dataset using a challenging species-agnostic approach that does not assume prior knowledge of species identity during outbreak detection. Performance was assessed using five metrics: normalized mutual information (NMI), adjusted rand index (ARI), sensitivity, PPV, and F1 score. To prevent distortion from singleton isolates, we calculated these metrics only for isolates forming ground truth clusters of at least two members. All metrics were averaged across four cross-validation folds. To ensure robustness, we conducted five independent trials with different random seeds controlling data splitting and model initialization. We reported the mean and the 95% confidence interval (using the *t*-distribution) of the averaged metrics across the five random trials.

**Table 2.** Species-agnostic clustering performance (mean ± 95% confidence interval) of evaluated alternatives to WGS-based clustering. The best result for each metric is highlighted in **bold**. Here, AR denotes antimicrobial resistance patterns; $\mathrm{MALDI}_{raw}$ refers to raw, binned MALDI-TOF spectra; $\mathrm{MALDI}_{ml}$ is the MALDI-TOF representation learned via our machine learning framework. $\mathrm{MALDI}_{sp}$ and $\mathrm{MALDI}_{snp}$ are ablation variants retaining only the species classification or SNP distance mapping objective, respectively. MALDI+AR is a fused representation of $\mathrm{MALDI}_{ml}$ and AR.

| **Model** | **NMI** | **ARI** | **Sensitivity** | **PPV** | **F1 score** |
|---|---|---|---|---|---|
| AR | 0.419 ± 0.006 | 0.009 ± 0.000 | **0.577 ± 0.003** | 0.018 ± 0.001 | 0.034 ± 0.001 |
| $\mathrm{MALDI}_{raw}$ | 0.679 ± 0.017 | 0.108 ± 0.012 | 0.205 ± 0.044 | 0.097 ± 0.008 | 0.123 ± 0.012 |
| $\mathrm{MALDI}_{ml}$ | 0.731 ± 0.017 | 0.160 ± 0.009 | 0.256 ± 0.059 | 0.143 ± 0.010 | 0.173 ± 0.009 |
| $\mathrm{MALDI}_{sp}$ | 0.716 ± 0.024 | 0.153 ± 0.006 | 0.275 ± 0.070 | 0.135 ± 0.012 | 0.167 ± 0.007 |
| $\mathrm{MALDI}_{snp}$ | 0.726 ± 0.017 | 0.107 ± 0.011 | 0.141 ± 0.033 | 0.114 ± 0.006 | 0.119 ± 0.012 |
| MALDI+AR | **0.783 ± 0.019** | **0.165 ± 0.022** | 0.209 ± 0.041 | **0.188 ± 0.024** | **0.175 ± 0.021** |

As shown in Table 2, the machine learning-based representations from $\mathrm{MALDI}_{ml}$ outperformed the $\mathrm{MALDI}_{raw}$ baseline across all metrics. This baseline used raw, binned MALDI-TOF spectra directly for clustering. These results indicate that our framework effectively extracts representations from MALDI-TOF spectra that are useful for outbreak detection.

To understand the contribution of each model component, we performed ablation studies. Removing either the species classification ($\mathrm{MALDI}_{snp}$) or the SNP distance mapping objective ($\mathrm{MALDI}_{sp}$) degraded performance in NMI, ARI, and PPV compared to the full model ($\mathrm{MALDI}_{ml}$), indicating that both objectives enhance outbreak detection capability. Notably, species classification alone ($\mathrm{MALDI}_{sp}$) achieved performance comparable to the full model and even yielded higher sensitivity. This finding suggests that disentangling species-level representations is essential for building a unified, multi-species framework, and that the model implicitly learns fine-grained, strain-level patterns that inform downstream clustering—even when trained only on species-level labels.

We further evaluated whether combining MALDI-TOF with antibiotic resistance (AR) data could improve performance. Consistent with prior studies, using AR data alone achieved the highest sensitivity but the lowest PPV among all methods. However, fusing MALDI and AR data produced the best overall results, outperforming either modality alone across PPV, NMI, and ARI. These findings demonstrate that MALDI-TOF and AR profiles capture complementary signals, and that integrating both modalities yields a more robust alternative to WGS-based gold standard for outbreak detection than AR or MALDI-TOF used separately.

*Species-specific utility*: We further examined the utility of MALDI-TOF, AR, and their combination in species-specific outbreak detection, where species identity is assumed to be known *a priori* and clustering is restricted to the isolates of each species.

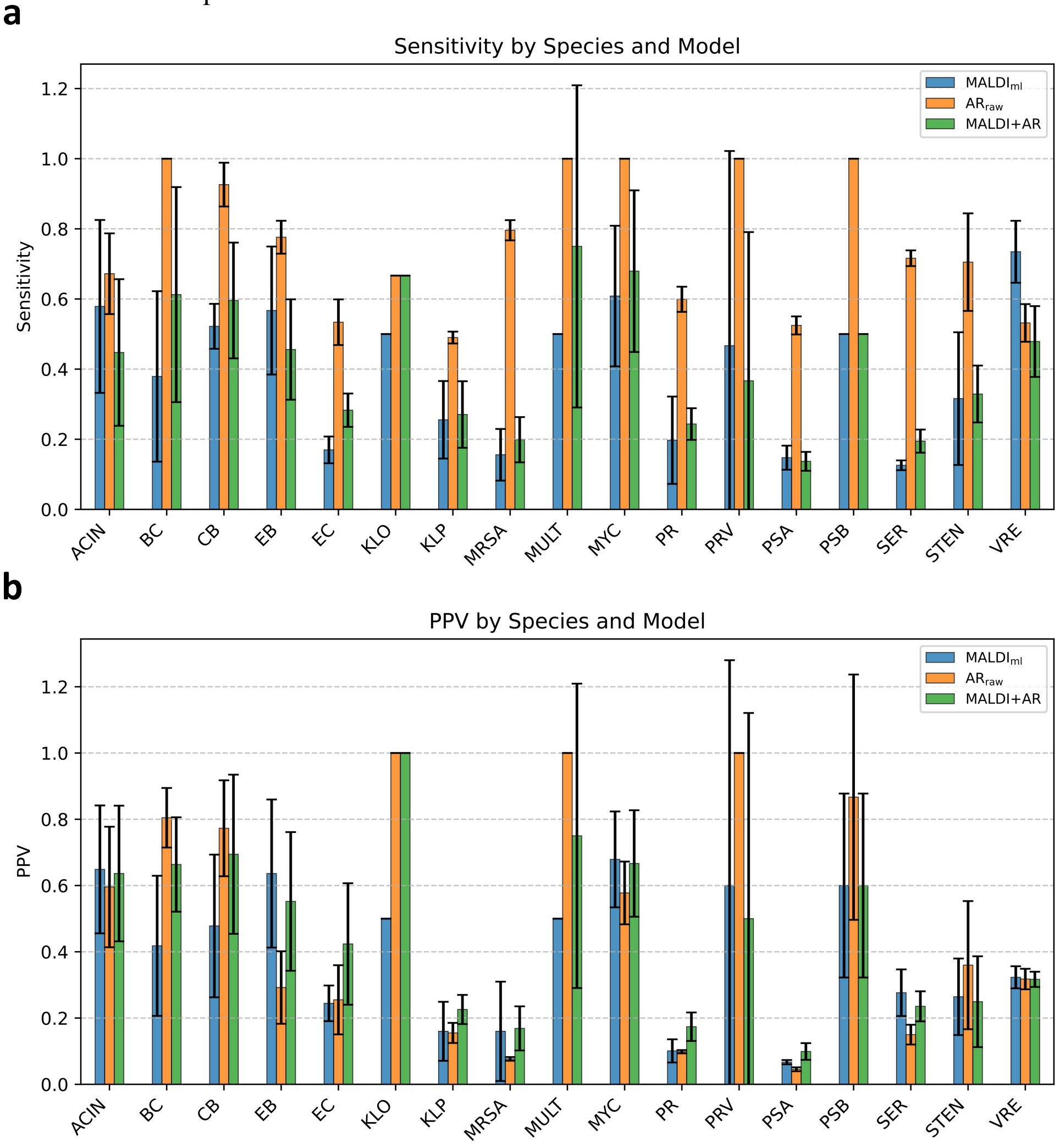

**Figure 2.** Per-species outbreak detection performance of MALDI$_{ml}$, AR$_{raw}$, and MALDI+AR in terms of sensitivity (**a**) and PPV (**b**).

We observed substantial variation in both sensitivity and PPV across species: sensitivity ranged from lower values in EC, KLP, PR, and PSA to higher values in ACIN, CB, MYC, and VRE (Figure 2a); similarly, PPV was lower in KLP, MRSA, PR, and PSA but higher in ACIN, CB, KLO, MYC, PRV, and PSB (Figure 2b).

We next compared the relative strengths of MALDI-TOF, AR, and their combination for species-specific outbreak detection. AR achieved superior specificity across all species except VRE, where MALDI-TOF substantially outperformed both AR alone and the MALDI+AR fusion approach (Figure 3a,b). Notably, fusing AR with MALDI-TOF increased specificity in several species, including BC, CB, EC, and KLO (Figure 3c), suggesting that AR provides key information that improves outbreak cluster identification.

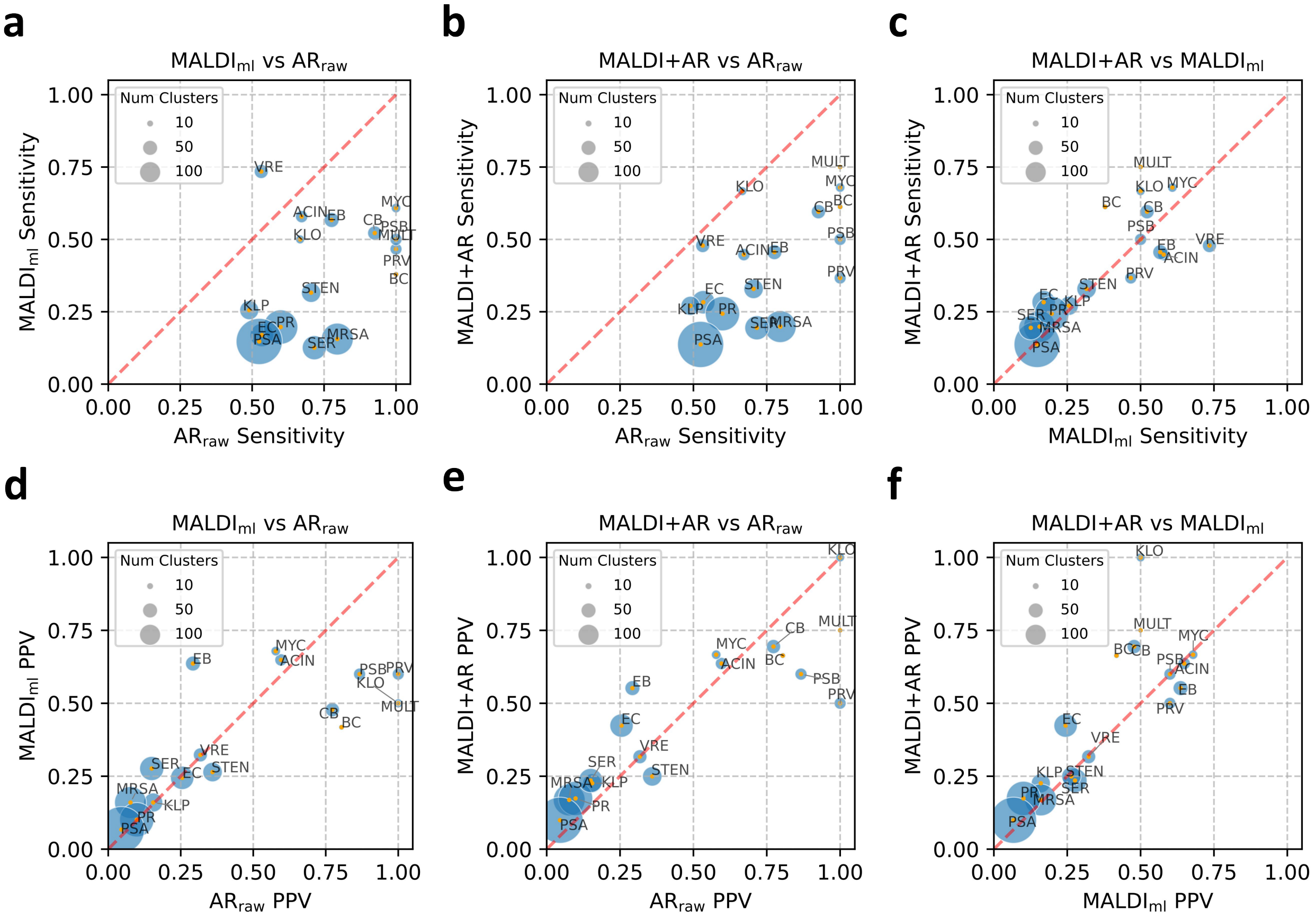

**Figure 3.** Scatterplots comparing the per-species sensitivity (**a-c**) and PPV (**d-f**) between MALDI$_{ml}$, AR$_{ml}$, and MALDI+AR. The sizes of the blue circles, with centers marked in orange, reflect the number of outbreak clusters in that species.

In terms of PPV, AR outperformed MALDI-TOF in multiple species, including BC, CB, KLO, PSB, PRV, and STEN (Figure 3d). However, combining both modalities yielded superior performance in EC, KLP, PSA, and PR compared to either modality alone (Figure 3e,f), indicating that MALDI-TOF and AR capture complementary information that enhances outbreak detection accuracy when used together.

Notably, across all metrics and methods, species with higher cluster complexity (e.g., PSA, PR, MRSA) consistently yielded lower performance, occupying the lower-left parts in Figure 3. This result highlights the challenge of detecting outbreaks in more diverse pathogen populations using these cost-effective alternatives.

In summary, MALDI-TOF and AR exhibited substantial species-specific variation in outbreak detection performance. For species-specific outbreak investigations, modality selection should be tailored to the pathogen species, leveraging their distinct and complementary strengths.

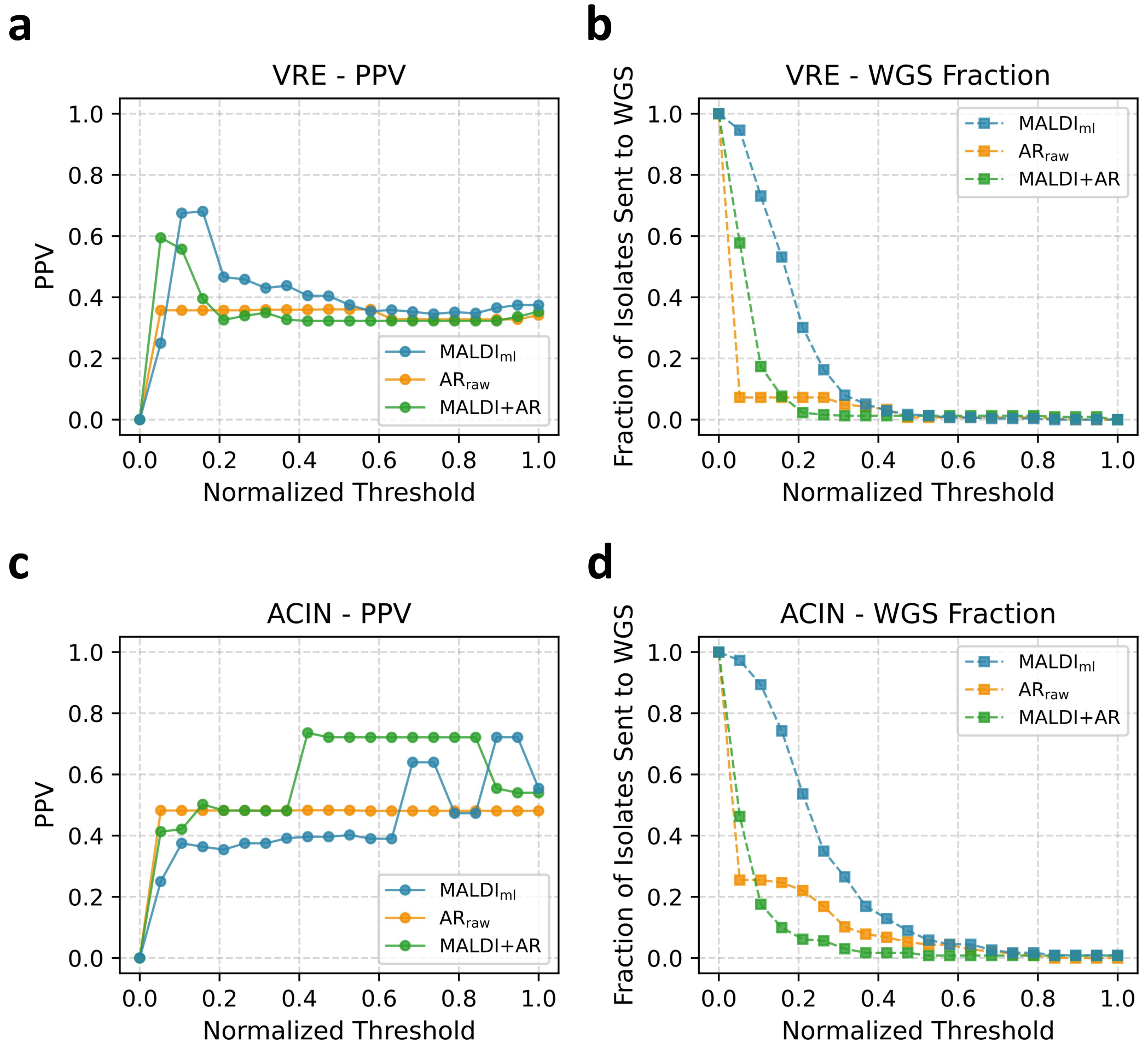

**Figure 4.** PPV (**a**,**c**) and fraction of isolates requiring WGS (**b**,**d**) at varying distance thresholds.

*Potential to reduce WGS dependency*: We further assessed the potential of MALDI-TOF and AR to reduce reliance on WGS for outbreak detection. We adopted a triage framework where isolates not assigned to any identified cluster are flagged for WGS confirmation. For each method among MALDI$_{ml}$, AR$_{raw}$, and MALDI+AR, we evaluated 20 equally spaced distance thresholds spanning the minimum to maximum Euclidean distances between its representati ons of the training data. Thresholds were normalized using min-max scaling to range from 0 to 1 for comparative visualization across methods and species.

Using vancomycin-resistant *Enterococcus faecium* (VRE) and *Acinetobacter baumannii* (ACIN) as representative examples, we examined how PPV (Figure 4a,c) and the fraction of isolates requiring WGS (Figure 4b,d) vary with distance threshold. As expected, increasing the threshold expanded cluster sizes and reduced the proportion of isolates referred to WGS; however, the corresponding PPV trends differed substantially between species.

For VRE, MALDI$_{ml}$ achieved the highest PPV at normalized thresholds below 0.2 (Figure 4a). However, more than 40% of isolates still required WGS at this operating point (Figure 4b). This observation indicates limited potential for MALDI-TOF or AR to substantially reduce WGS dependency in VRE. In contrast, for ACIN, MALDI+AR achieved the highest PPV at normalized thresholds between 0.4 and 0.8 (Figure 4c) while reducing the fraction of isolates requiring WGS to nearly zero (Figure 4d). This result demonstrates that the combined modality can dramatically reduce WGS reliance for ACIN while maintaining high precision in outbreak detection. These contrasting behaviors illustrate that the extent to which MALDI-TOF and AR can reduce dependency on WGS is species-dependent, and that species-specific calibration of distance thresholds is essential to identify operating regimes that balance precision against downstream sequencing burden.

*Simulated operational cost analysis*: Since the practical utility of MALDI-TOF and AR is strongly influenced by cost and user accessibility considerations, we conducted a simulated operational cost analysis using the same decision framework, in which isolates not assigned to a cluster are subjected to WGS. We applied previously reported cost estimates for WGS[30], MALDI-TOF[31,32], and AR[33]. We assigned a cost of 100 abstract units per isolate to WGS and s imulated two scenarios: (1) MALDI-TOF (1 unit) cheaper than AR (5 units), and (2) AR (1 unit) cheaper than MALDI-TOF (5 units). Note that this analysis focuses on direct procedural costs and does not account for downstream costs associated with false-positive or false-negative cluster assignments, which depend on complex temporal and institution-specific factors beyond the scope of this study.

Using ACIN as an illustrative case, we confirmed that increasing the distance threshold reduced total operational cost in both scenarios (Figure 5a,b), consistent with the corresponding reduction in WGS utilization (Figure 4d). More importantly, when operational cost was plotted against PPV, both $MALDI_{ml}$ and MALDI+AR achieved high PPV at relatively low normalized cost, occupying the favorable lower-right quadrant of the cost–performance space. On the other hand, AR failed to achieve comparably high PPV across the explored thresholds (Figure 5c,d). This demonstrates that MALDI-TOF-based approaches enable cost-effective yet relatively accurate outbreak surveillance for ACIN. These cost-performance profiles can provide practitioners with quantitative guidance for selecting modalities and distance thresholds that balance accuracy with resource constraints under local cost structures.

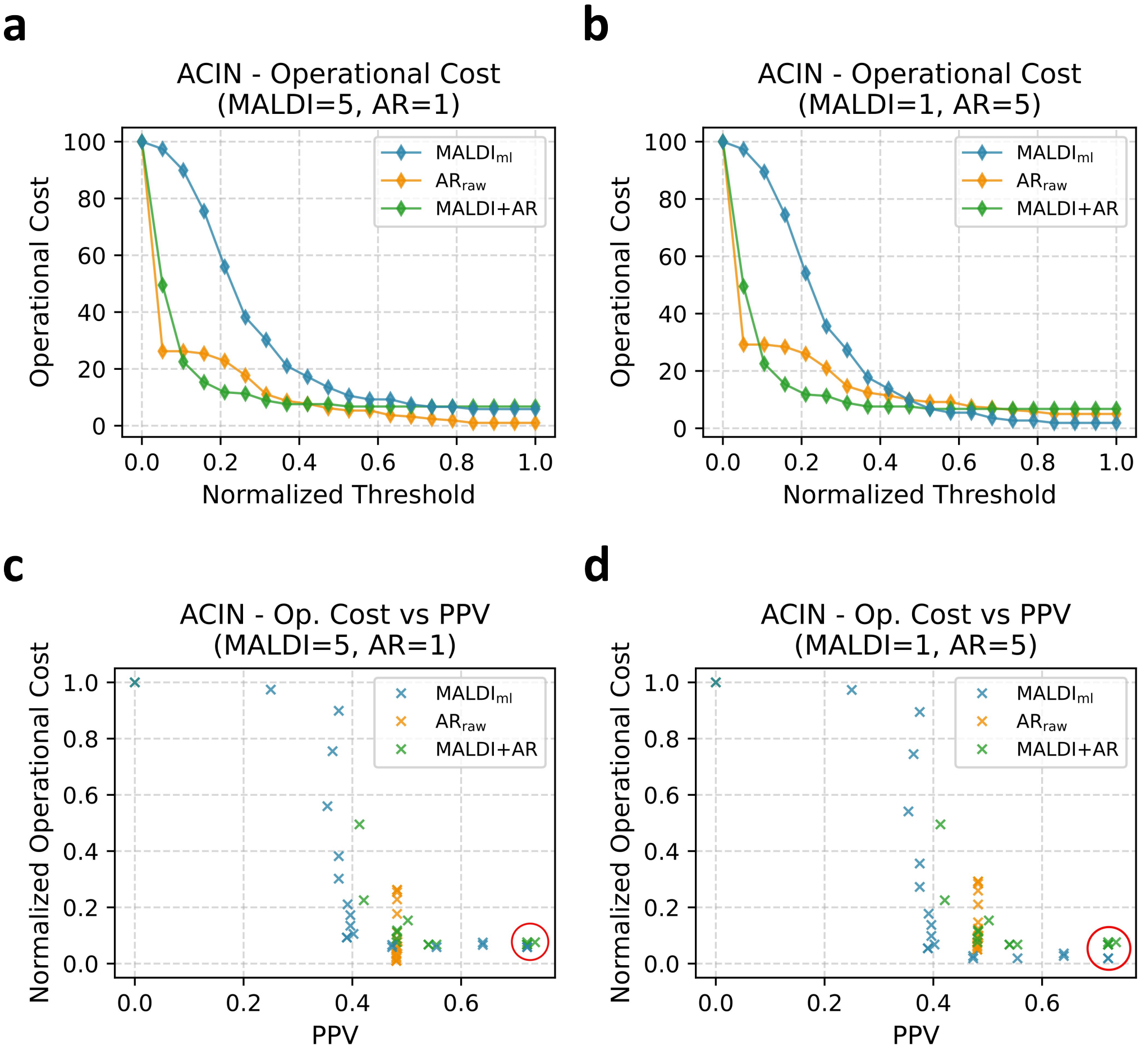

**Figure 5.** Simulated operational costs at varying distance thresholds under alternative cost scenarios. WGS cost is set to 100 units per isolate. **a, c:** MALDI-TOF = 5 units, AR = 1 unit. **b, d:** MALDI-TOF = 1 unit, AR = 5 units. The no rmalized operational cost is computed by deriving the actual cost by 100, the maximum cost. The samples occupyin g the favorable lower-right quadrant of the cost–performance space are circled in red.

## Discussion

In this study, we developed and evaluated a machine learning framework for hospital outbreak detection using MALDI-TOF mass spectrometry and antimicrobial resistance (AR) profiles as rapid, low-cost alternatives to whole-genome sequencing (WGS). By leveraging phenotypic data already generated in routine clinical microbiology workflows, the proposed approaches aim to enable rapid yet more scalable outbreak detection while reserving WGS for confirmatory and high-resolution analyses.

While integrating both MALDI-TOF and AR yielded the best results in species-agnostic settings, species-specific analyses revealed substantial heterogeneity in performance, indicating that no single modality is universally plausibl e. AR profiles generally provided higher specificity, except in VRE, where MALDI-TOF outperforming AR suggests that resistance patterns alone may be insufficient to resolve transmission chains in certain species. Importantly, integrating MALDI-TOF with AR consistently improved performance in several challenging species, including EC, KLP, PSA, and PR, highlighting the complementary nature of spectral and resistance-derived information. These species-specific patterns likely reflect underlying differences in resistance biology and population genetics, and understanding these determinants represents an important direction for optimizing phenotypic surveillance strategies.

Our results further illustrate that the ability of rapid phenotypic methods to reduce dependence on WGS is strongly pathogen-dependent. For some organisms, such as ACIN, MALDI-TOF and AR—particularly when combined—can substantially reduce the need for routine sequencing while maintaining high precision. For others, including VRE, phenotypic similarity alone is insufficient to confidently exclude relatedness, necessitating continued reliance on WGS for accurate outbreak confirmation. These findings support a tiered, species-aware surveillance paradigm, in which MALDI-TOF and AR function as front-line triage tools that prioritize isolates for genomic sequencing rather than replacing WGS outright. For instance, isolates belonging to species that achieve high PPV with MALDI-TOF and AR require minimal downstream WGS, while those with low PPV warrant heightened use of WGS to maintain outbreak detection precision. More work however is needed to assess practical utility of a hybrid approach that would apply probabilistic modeling to decide based on MALDI-TOF and/or AR signatures and the information already available for a developing cluster which new isolates to assign for WGS fingerprinting.

To evaluate practical deployment considerations, we conducted a simple simulated operational cost analysis that explicitly incorporates downstream WGS utilization. The analysis demonstrates how different modality choices and clustering thresholds translate into distinct PPV–cost trade-offs and provides a quantitative basis for selecting operating points under varying local resource constraints. While the cost assumptions are illustrative, the framework is adaptable to institution-specific pricing and laboratory workflows, aligning with real-world decision-making needs in hospital environments. However, more work is needed to include time value of information into consideration. E.g., in rural hospital settings where WGS capability is not available on premises but can be done offsite, obtaining sequencing data for isolates may add more latency to clustering process than in flagship clinical facilities, impacting timeliness of outbreak detection and the ability of infection control teams to mitigate emerging crises. These kinds of tradeoffs could change the relative utility of alternatives to genome sequencing in their favor in some operational settings.

Several limitations warrant consideration. Our cost analysis focuses on direct operational costs and does not capture the downstream impact of outbreak detection errors. False-positive cluster assignments may prompt unnecessary infection control interventions, while false negatives may allow continued transmission and delayed responses. Accurately modeling these error-related costs requires accounting for temporal dynamics, patient movement, and intervention effects, which were beyond the scope of this study. Future work will focus on integrating such factors to better quantify the clinical and operational consequences of outbreak detection errors.

Finally, while this work centers on microbiology-derived features, outbreak detection can benefit from incorporating additional data sources. Electronic health records (EHRs) contain rich contextual information on patient location, movement, and care processes that can inform transmission pathways[34]. Future efforts will explore extracting and integrating EHR-derived representations with MALDI-TOF and AR profiles to build more comprehensive, multimodal outbreak surveillance systems.

## Acknowledgements

This work was funded in part by the National Institute of Allergy and Infectious Diseases, National Institutes of Health (NIH) grant R01AI127472 and by the National Science Foundation award 2406231.

## References

1. Epstein L, Hunter JC, Arwady MA, Tsai V, Stein L, Gribogiannis M, Frias M, Guh AY, Laufer AS, Black S, Pacilli M. New Delhi metallo-β-lactamase–producing carbapenem-resistant Escherichia coli associated with exposure to duodenoscopes. Jama. 2014 Oct 8;312(14):1447-55.
2. Danzmann L, Gastmeier P, Schwab F, Vonberg RP. Health care workers causing large nosocomial outbreaks: a systematic review. BMC infectious diseases. 2013 Feb 22;13(1):98.Gardner RM, Golubjatnikov OK, Laub RM, Jacobson JT, Evans RS. Computer-critiqued blood ordering using the HELP system. Comput Biomed Res 1990;23:514-28.
3. Palmore TN, Henderson DK. Managing transmission of carbapenem-resistant enterobacteriaceae in healthcare settings: a view from the trenches. Clinical Infectious Diseases. 2013 Dec 1;57(11):1593-9.
4. Archibald LK, Jarvis WR. Health care–associated infection outbreak investigations by the Centers for Disease Control and Prevention, 1946–2005. American journal of epidemiology. 2011 Dec 1;174(suppl_11):S47-64.
5. Poon AF. Impacts and shortcomings of genetic clustering methods for infectious disease outbreaks. Virus evolution. 2016 Jul;2(2):vew031.
6. Quainoo S, Coolen JP, van Hijum SA, Huynen MA, Melchers WJ, van Schaik W, Wertheim HF. Whole-genome sequencing of bacterial pathogens: the future of nosocomial outbreak analysis. Clinical microbiology reviews. 2017 Oct;30(4):1015-63.
7. Foxman B, Riley L. Molecular epidemiology: focus on infection. American journal of epidemiology. 2001 Jun 15;153(12):1135-41.
8. Mellmann A, Bletz S, Böking T, Kipp F, Becker K, Schultes A, Prior K, Harmsen D. Real-time genome sequencing of resistant bacteria provides precision infection control in an institutional setting. Journal of clinical microbiology. 2016 Dec;54(12):2874-81.
9. Quick J, Cumley N, Wearn CM, Niebel M, Constantinidou C, Thomas CM, Pallen MJ, Moiemen NS, Bamford A, Oppenheim B, Loman NJ. Seeking the source of Pseudomonas aeruginosa infections in a recently opened hospital: an observational study using whole-genome sequencing. BMJ open. 2014 Nov 1;4(11):e006278.
10. Sundermann AJ, Chen J, Miller JK, Saul MI, Shutt KA, Griffith MP, Mustapha MM, Ezeonwuka C, Waggle K, Srinivasa V, Kumar P. Outbreak of Pseudomonas aeruginosa infections from a contaminated gastroscope detected by whole genome sequencing surveillance. Clinical Infectious Diseases. 2021 Aug 1;73(3):e638-42.
11. Abdelbary MH, Senn L, Greub G, Chaillou G, Moulin E, Blanc DS. Whole-genome sequencing revealed independent emergence of vancomycin-resistant Enterococcus faecium causing sequential outbreaks over 3 years in a tertiary care hospital. European Journal of Clinical Microbiology & Infectious Diseases. 2019 Jun 1;38(6):1163-70.
12. Sundermann AJ, Babiker A, Marsh JW, Shutt KA, Mustapha MM, Pasculle AW, Ezeonwuka C, Saul MI, Pacey MP, Van Tyne D, Ayres AM. Outbreak of vancomycin-resistant Enterococcus faecium in interventional radiology: detection through whole-genome sequencing-based surveillance. Clinical Infectious Diseases. 2020 May 23;70(11):2336-43.
13. Stimson J, Gardy J, Mathema B, Crudu V, Cohen T, Colijn C. Beyond the SNP threshold: identifying outbreak clusters using inferred transmissions. Molecular biology and evolution. 2019 Mar 1;36(3):587-603.
14. Hatherell HA, Colijn C, Stagg HR, Jackson C, Winter JR, Abubakar I. Interpreting whole genome sequencing for investigating tuberculosis transmission: a systematic review. BMC medicine. 2016 Mar 23;14(1):21.
15. Rossen JW, Friedrich AW, Moran-Gilad J. Practical issues in implementing whole-genome-sequencing in routine diagnostic microbiology. Clinical microbiology and infection. 2018 Apr 1;24(4):355-60.
16. Croxatto A, Prod'hom G, Greub G. Applications of MALDI-TOF mass spectrometry in clinical diagnostic microbiology. FEMS microbiology reviews. 2012 Mar 1;36(2):380-407.
17. Clark AE, Kaleta EJ, Arora A, Wolk DM. Matrix-assisted laser desorption ionization–time of flight mass spectrometry: a fundamental shift in the routine practice of clinical microbiology. Clinical microbiology reviews. 2013 Jul;26(3):547-603.
18. Griffin PM, Price GR, Schooneveldt JM, Schlebusch S, Tilse MH, Urbanski T, Hamilton B, Venter D. Use of matrix-assisted laser desorption ionization–time of flight mass spectrometry to identify vancomycin-resistant enterococci and investigate the epidemiology of an outbreak. Journal of clinical microbiology. 2012 Sep;50(9):2918-31.
19. Rödel J, Mellmann A, Stein C, Alexi M, Kipp F, Edel B, Dawczynski K, Brandt C, Seidel L, Pfister W, Löffler B. Use of MALDI-TOF mass spectrometry to detect nosocomial outbreaks of Serratia marcescens and Citrobacter freundii. European Journal of Clinical Microbiology & Infectious Diseases. 2019 Mar 4;38(3):581-91.

20. Giraud-Gatineau A, Texier G, Fournier PE, Raoult D, Chaudet H. Using MALDI-TOF spectra in epidemiological surveillance for the detection of bacterial subgroups with a possible epidemic potential. BMC Infectious Diseases. 2021 Oct 28;21(1):1109.
21. Williamson YM, Moura H, Woolfitt AR, Pirkle JL, Barr JR, Carvalho MD, Ades EP, Carlone GM, Sampson JS. Differentiation of Streptococcus pneumoniae conjunctivitis outbreak isolates by matrix-assisted laser desorption ionization-time of flight mass spectrometry. Applied and environmental microbiology. 2008 Oct 1;74(19):5891-7.
22. Schlebusch S, Price GR, Hinds S, Nourse C, Schooneveldt JM, Tilse MH, Liley HG, Wallis T, Bowling F, Venter D, Nimmo GR. First outbreak of PVL-positive nonmultiresistant MRSA in a neonatal ICU in Australia: comparison of MALDI-TOF and SNP-plus-binary gene typing. European journal of clinical microbiology & infectious diseases. 2010 Oct;29(10):1311-4.
23. Schlebusch S, Price GR, Gallagher RL, Horton-Szar V, Elbourne LD, Griffin P, Venter DJ, Jensen SO, Van Hal SJ. MALDI-TOF MS meets WGS in a VRE outbreak investigation. European Journal of Clinical Microbiology & Infectious Diseases. 2017 Mar;36(3):495-9.
24. Dinkelacker AG, Vogt S, Oberhettinger P, Mauder N, Rau J, Kostrzewa M, Rossen JW, Autenrieth IB, Peter S, Liese J. Typing and species identification of clinical Klebsiella isolates by Fourier transform infrared spectroscopy and matrix-assisted laser desorption ionization–time of flight mass spectrometry. Journal of clinical microbiology. 2018 Nov;56(11):10-128.
25. Stelling J, Yih WK, Galas M, Kulldorff M, Pichel M, Terragno R, Tuduri E, Espetxe S, Binsztein N, O'BRIEN TF, Platt R. Automated use of WHONET and SaTScan to detect outbreaks of Shigella spp. using antimicrobial resistance phenotypes. Epidemiology & Infection. 2010 Jun;138(6):873-83.
26. Natale A, Stelling J, Meledandri M, Messenger LA, D'Ancona F. Use of WHONET-SaTScan system for simulated real-time detection of antimicrobial resistance clusters in a hospital in Italy, 2012 to 2014. Eurosurveillance. 2017 Mar 16;22(11):30484.
27. Hosaka Y, Hirabayashi A, Clark A, Baker M, Sugai M, Stelling J, Yahara K. Enhanced automated detection of outbreaks of a rare antimicrobial-resistant bacterial species. Plos one. 2024 Oct 24;19(10):e0312477.
28. Stachel A, Pinto G, Stelling J, Shopsin B, Inglima K, Phillips M. An Evaluation of an Automated Hospital Outbreak Detection System (WHONET-SaTScan) Versus Standard Outbreak Detection Approach. In Open Forum Infectious Diseases 2016 Dec 1 (Vol. 3, No. suppl_1, p. 1365). Oxford University Press.
29. Liu C, Chen J, Sundermann AJ, Harrison LH, Dubrawski A. Bridging the utility gap between MALDI-TOF and WGS for affordable outbreak cluster detection. Proceedings of Machine Learning Research. 2025;287:1-5.
30. Waggle KD, Griffith MP, Rokes AB, Srinivasa VR, Nawrocki EM, Ereifej D, Patrick R, Coyle H, Chaudhary S, Raabe NJ, Shutt K. Methods for cost-efficient, whole genome sequencing surveillance for enhanced detection of outbreaks in a hospital setting. medRxiv. 2025 Oct 17:2024-02.
31. Dhiman N, Hall L, Wohlfiel SL, Buckwalter SP, Wengenack NL. Performance and cost analysis of matrix-assisted laser desorption ionization–time of flight mass spectrometry for routine identification of yeast. Journal of clinical microbiology. 2011 Apr;49(4):1614-6.
32. Gaillot O, Blondiaux N, Loïez C, Wallet F, Lemaître N, Herwegh S, Courcol RJ. Cost-effectiveness of switch to matrix-assisted laser desorption ionization-time of flight mass spectrometry for routine bacterial identification. Journal of clinical microbiology. 2011 Dec;49(12):4412-.
33. Turner GP, Dusich I, Thomson RB Jr. Comparative cost of antimicrobial susceptibility testing using BIOMIC disk diffusion vs. the BD Phoenix™ AP automated microbiology systems. Poster presented at: American Society for Microbiology General Meeting; 2010 May 23-27; San Diego, CA.
34. Miller JK, Chen J, Sundermann A, Marsh JW, Saul MI, Shutt KA, Pacey M, Mustapha MM, Harrison LH, Dubrawski A. Statistical outbreak detection by joining medical records and pathogen similarity. Journal of biomedical informatics. 2019 Mar 1;91:103126.